\documentclass[12pt]{article}
\usepackage{amsmath}
\usepackage{cancel}

\newcommand{\Msusy}{M_{\textrm{SUSY}}}

\def\beq{\begin{equation}}
\def\eeq#1{\label{#1}\end{equation}}
\def\eeqn{\end{equation}}

\def\beqa{\begin{eqnarray}}
\def\eeqa#1{\label{#1}\end{eqnarray}}
\def\eeqan{\end{eqnarray}}

\let\bar=\overbar

\def\Dslash{\not{\hbox{\kern-4pt $D$}}}
\def\dslash{\not{\hbox{\kern-2pt $\del$}}}

\def\msb{{\bar{\ssstyle M \kern -1pt S}}}

\def\Title#1{\begin{center} {\Large {\bf #1} } \end{center}}

\begin{document}

\Title{Chirally enhanced corrections in the MSSM\footnote{Work done in collaboration with Lars Hofer and Janusz Rosiek}}

\bigskip\bigskip

%+\addtocontents{toc}{{\it D. Reggiano}}
%+\label{ReggianoStart}

\begin{raggedright}  

{\it Andreas Crivellin \index{Crivellin, A.}\\
Albert Einstein Center for Fundamental Physics\\
Institute for Theoretical Physics\\
CH-3012 Bern, Switzerland}
\bigskip\bigskip
\end{raggedright}

In the general MSSM, chirally-enhanced corrections (to Yukawa couplings) are induced by
gluino-squark, chargino-sfermion and neutralino-sfermion loops and can
numerically compete with, or even dominate over, tree-level
contributions, due to their enhancement by either $\tan\beta$ or
$A^f_{ij}/(Y^f_{ij}\Msusy)$.  Building up on earlier work \cite{Crivellin:2008mq} we identified all
potential chirally-enhanced corrections (flavor-conserving and flavor-changing) in the generic MSSM in Ref.~\cite{chiral} and discussed their
effects on the finite renormalization of Yukawa couplings, fermion
wave-functions and the CKM matrix. To leading order in $v/\Msusy$ (the so-called decoupling limit),
which numerically is a very good approximation for realistic choices
of MSSM parameters, we obtained analytic resummation formulae for
these quantities. 

For relating fermion masses $m_{f_i}$ to their Yukawa couplings $Y^{f_i(0)}$ the procedure is rather straightforward. After calculating all chirally enhanced pieces of the fermion self-energies one has to identify which parts of the self-energies depend on Yukawa couplings and which are independent of them. In the decoupling limit the self-energies can at most involve one power of $Y^{f(0)}$. For definiteness we consider down-quarks here and  decompose the self-energy as follows: 
\begin{equation}
\Sigma_{ii}^{d\,LR} \;=\;
\Sigma_{ii\,\cancel{Y_i}}^{d\,LR} \, + \,
\epsilon_i^{d}\,v_u\,\,Y^{d_i(0)}\,.
\label{eq:epsilon_b}
\end{equation}
Now the bare Yukawa coupling can be calculated in terms of the physical one by solving $v_d Y^{d_i(0)}+\Sigma_{ii}^{d\,LR}=m_{d_i}$ which gives
\begin{equation}
Y^{d_i(0)} = \dfrac{m_{d_i} - \Sigma_{ii\,\cancel{Y_i}}^{d\,LR}}{v_d
  \left( {1 + \tan\beta \varepsilon_i^d } \right)}.
\label{md-Yd}
\end{equation}
For the CKM resummation the procedure is more complicated. Here a rotation in flavor space of the quark fields is induced by the self-energy corrections
\begin{equation}
\renewcommand{\arraystretch}{2.2}
U_{fi}^{q\;L}  = \left( {\begin{array}{*{20}c}
   1 & {\dfrac{{\Sigma _{12}^{q\;LR} }}{{m_{q_2 } }}} & {\dfrac{{\Sigma _{13}^{q\;LR} }}{{m_{q_3 } }}}  \\
   {\dfrac{{ - \Sigma _{21}^{q\;RL} }}{{m_{q_2 } }}} & 1 & {\dfrac{{\Sigma _{23}^{q\;LR} }}{{m_{q_3 } }}}  \\
   {\dfrac{{ - \Sigma _{31}^{q\;RL} }}{{m_{q_3 } }}}+{\dfrac{{  \Sigma _{32}^{q\;RL}\Sigma _{21}^{q\;RL} }}{{m_{q_3 }m_{q_2 } }}} & {\dfrac{{ - \Sigma _{32}^{q\;RL} }}{{m_{q_3 } }}} & 1  \\
\end{array}} \right)_{fi}.
\label{DeltaU}
\end{equation}
For explicitly solving the equation
\begin{equation}
V^{CKM\left( 0 \right)}  = U^{u\;L\dag } V^{CKM} U^{d\;L} 
\label{CKM-0-ren}
\end{equation}
for the bare CKM matrix it is necessary to divide the self-energies into parts proportional to CKM elements and parts independent of them and to exploit in addition the CKM hierarchy. Because the bare CKM matrix (to order $\lambda^3$) has three phases, one has to extend the classical Wolfenstein parametrization allowing for complex $\lambda$ and $A$ parameters (given in the appendix of Ref.~\cite{chiral}).  

Knowing the bare CKM matrix and the bare Yukawa couplings the chirally enhanced corrections can be absorbed into effective vertices. In principle, the procedure for fermion-sfermion-gaugino(higgsino) vertices is simple: One has to insert the bare quantities into the expressions for the Feynman rules and in addition apply the rotations in flavor space $U^{f\,L,R}$ to all SM fermion lines involved in the vertex. If these effective Feynman rules are used for the calculation of an Feynman amplitude at leading order in perturbation theory, all kinds
of chirally-enhanced effects are automatically included and resummed
to all orders in the final result.

For the calculation of effective Higgs-fermion-fermion vertices also a decomposition of the self-energies into holomorphic and non-holomorphic parts is necessary:
\begin{equation}
\Sigma_{ji}^{f\,LR} = \Sigma_{ji\,A}^{f\,LR} + \Sigma_{ji}^{\prime f\,LR}\,.  \label{HoloDeco}
\end{equation}
Here the holomorphic part $\Sigma_{ji\,A}^{f\,LR}$ is proportional to an $A$-term while the chirality flip in the non-holomorphic part $\Sigma_{ji}^{\prime f\,LR}$ is provided by a Yukawa coupling (or an $A^{\prime}$ term). The Higgs vertices are most easily calculated in the effective 2HDM obtained after integrating out the SUSY particles. For example the coupling of down-quarks to the CP-odd Higgs is then (in the large $\tan\beta$ limit) given by:
\begin{equation}
{\Gamma_{d_f d_i }^{H_k^0\,LR\,\rm{eff} } } = \dfrac{i}{\sqrt{2 }v}\tan\beta \left( m_{d_i} \delta_{fi} - \widetilde \Sigma_{fi}^{\prime d} \right)\;\;\;{\rm with}\;\;\;\widetilde{\Sigma}^{\prime d\,LR}_{fi}=U_{jf}^{d\,L*} \Sigma^{\prime d\,LR}_{jk}
U_{ki}^{d\,R}
 \label{Higgs-vertices-decoupling}
\end{equation} 
Note that in this way also the $A$-terms can lead to flavor-changing neutral Higgs couplings at large $\tan\beta$. These effective Higgs-vertices can be used in the limit
$m_{H^0},m_{A^0},m_{H^{\pm}}\ll \Msusy$ as Feynman rules in an
effective theory with the SUSY particles being integrated
out. However, they remain valid in the case
$m_{H^0},m_{A^0},m_{H^{\pm}} \sim \Msusy$ as long as the momenta
flowing through the Higgs vertices are much smaller than $\Msusy$.
Thus our effective Higgs-fermion-fermion Feynman rules can e.g. be
applied to calculate Higgs penguins contributing to
$B_{d,s}\to\mu^+\mu^-$, $B^+\to\tau^+\nu$ or the double Higgs penguin
contributing to $\Delta F=2$ processes.


\begin{thebibliography}{99}


  %\cite{chiral}
  
  %\cite{Crivellin:2008mq}
\bibitem{Crivellin:2008mq}
%\cite{Crivellin:2010er}
%\bibitem{Crivellin:2010er}
  A.~Crivellin,
  %``Effective Higgs Vertices in the generic MSSM,''
  Phys.\ Rev.\  D {\bf 83} (2011) 056001
  [arXiv:1012.4840 [hep-ph]].
  %%CITATION = PHRVA,D83,056001;%%
  A.~Crivellin and U.~Nierste,
  %``Supersymmetric renormalisation of the CKM matrix and new constraints on the
  %squark mass matrices,''
  Phys.\ Rev.\  D {\bf 79} (2009) 035018
  [arXiv:0810.1613 [hep-ph]].
  %%CITATION = PHRVA,D79,035018;%%
%\cite{Crivellin:2009ar}
%\bibitem{Crivellin:2009ar}
  A.~Crivellin and U.~Nierste,
  %``Chirally enhanced corrections to FCNC processes in the generic MSSM,''
  Phys.\ Rev.\  D {\bf 81} (2010) 095007
  [arXiv:0908.4404 [hep-ph]].
  %%CITATION = PHRVA,D81,095007;%%
%\cite{Crivellin:2010gw}
%\bibitem{Crivellin:2010gw}
  A.~Crivellin and J.~Girrbach,
  %``Constraining the MSSM sfermion mass matrices with light fermion masses,''
  Phys.\ Rev.\  D {\bf 81} (2010) 076001
  [arXiv:1002.0227 [hep-ph]].
  %%CITATION = PHRVA,D81,076001;%%
  %\cite{Hofer:2009xb}
%\bibitem{Hofer:2009xb}
  L.~Hofer, U.~Nierste and D.~Scherer,
  %``Resummation of tan-beta-enhanced supersymmetric loop corrections beyond the
  %decoupling limit,''
  JHEP {\bf 0910} (2009) 081
  [arXiv:0907.5408 [hep-ph]].
  %%CITATION = JHEPA,0910,081;%%
  %\cite{Girrbach:2009uy}
%\bibitem{Girrbach:2009uy}
  J.~Girrbach, S.~Mertens, U.~Nierste and S.~Wiesenfeldt,
  %``Lepton flavour violation in the MSSM,''
  JHEP {\bf 1005} (2010) 026
  [arXiv:0910.2663 [hep-ph]].
  %%CITATION = JHEPA,1005,026;%%
  
\bibitem{chiral}

  A.~Crivellin, L.~Hofer and J.~Rosiek,
  %``Complete resummation of chirally-enhanced loop-effects in the MSSM with
  %non-minimal sources of flavor-violation,''
  arXiv:1103.4272 [hep-ph].
  %%CITATION = ARXIV:1103.4272;%%
  


%  %\cite{Crivellin:2010ty}
%\bibitem{Crivellin:2010ty}
% A.~Crivellin, J.~Girrbach and U.~Nierste,
  %``Yukawa coupling and anomalous magnetic moment of the muon: an update for
  %the LHC era,''
%  Phys.\ Rev.\  D {\bf 83} (2011) 055009
%  [arXiv:1010.4485 [hep-ph]].
  %%CITATION = PHRVA,D83,055009;%%
 %\bibitem{Crivellin:2011XX}
%  A.~Crivellin, L.~Hofer, U.~Nierste and D.~Scherer,
  %``Phenomenological concequences of radiative flavor violation in the MSSM,''
%    [arXiv:1105.2818 [hep-ph]].



%\cite{Hall:1993gn}
%\bibitem{Hall:1993gn}
%  L.~J.~Hall, R.~Rattazzi and U.~Sarid,
  %``The Top quark mass in supersymmetric SO(10) unification,''
%  Phys.\ Rev.\  D {\bf 50} (1994) 7048
%  [arXiv:hep-ph/9306309].
  %%CITATION = PHRVA,D50,7048;%%
  
    %\cite{Buchmuller:1982ye}
%\bibitem{Buchmuller:1982ye}
%  W.~Buchmuller and D.~Wyler,
  %``CP Violation and R Invariance in Supersymmetric Models of Strong and
  %Electroweak Interactions,''
%  Phys.\ Lett.\  B {\bf 121} (1983) 321.
  %%CITATION = PHLTA,B121,321;%%
  %\cite{Borzumati:1999sp}
%\bibitem{Borzumati:1999sp}
%  F.~Borzumati, G.~R.~Farrar, N.~Polonsky and S.~D.~Thomas,
  %``Soft Yukawa couplings in supersymmetric theories,''
%  Nucl.\ Phys.\  B {\bf 555} (1999) 53
%  [arXiv:hep-ph/9902443].
  %%CITATION = NUPHA,B555,53;%%
  
    %\cite{Carena:1999py}
%\bibitem{Carena:1999py}
%  M.~S.~Carena, D.~Garcia, U.~Nierste and C.~E.~M.~Wagner,
  %``Effective Lagrangian for the $\bar{t} b H^{+}$ interaction in the MSSM and
  %charged Higgs phenomenology,''
%  Nucl.\ Phys.\  B {\bf 577} (2000) 88
%  [arXiv:hep-ph/9912516].
  %%CITATION = NUPHA,B577,88;%%
  
  %\cite{Guasch:2003cv}
%\bibitem{Guasch:2003cv}
%  J.~Guasch, P.~Hafliger and M.~Spira,
  %``MSSM Higgs decays to bottom quark pairs revisited,''
%  Phys.\ Rev.\  D {\bf 68} (2003) 115001
%  [arXiv:hep-ph/0305101].
    %%CITATION = PHRVA,D68,115001;%%
  
  %\cite{Casas:1995pd}
%\bibitem{Casas:1995pd}
%  J.~A.~Casas, A.~Lleyda and C.~Munoz,
  %``Strong constraints on the parameter space of the MSSM from charge and color
  %breaking minima,''
%  Nucl.\ Phys.\  B {\bf 471} (1996) 3
%  [arXiv:hep-ph/9507294].
  %%CITATION = NUPHA,B471,3;%%
  %\cite{Park:2010wf}
%\bibitem{Park:2010wf}
%  J.~h.~Park,
  %``Metastability bounds on flavour-violating trilinear soft terms in the
  %MSSM,''
%  Phys.\ Rev.\  D {\bf 83} (2011) 055015
%  [arXiv:1011.4939 [hep-ph]].
  %%CITATION = PHRVA,D83,055015;%%


\end{thebibliography}
\end{document}